\documentclass[11pt]{article}
\usepackage{graphicx}

\def\singleandabitspaced{\baselineskip=\normalbaselineskip\multiply
    \baselineskip by 110\divide\baselineskip by 100}
\def\singlespaced{\baselineskip=\normalbaselineskip}

\newcommand{\centeron}[2]{{\setbox0=\hbox{#1}\setbox1=\hbox{#2}\ifdim
                             \wd1>\wd0\kern.5\wd1\kern-.5\wd0\fi \copy0
                             \kern-.5\wd0\kern-.5\wd1\copy1\ifdim\wd0>\wd1
                             \kern.5\wd0\kern-.5\wd1\fi}}
\newcommand{\ltap}{\>\centeron{\raise.35ex\hbox{$<$}}
                     {\lower.65ex\hbox{$\sim$}}\>}
\newcommand{\gtap}{\>\centeron{\raise.35ex\hbox{$>$}}
                     {\lower.65ex\hbox{$\sim$}}\>}

\begin{document}

\singlespaced

\begin{titlepage}

\begin{center}
\vspace*{0.8in} \mbox{\Large \textbf{Have Atmospheric \u{C}erenkov
                                     Telescopes Observed Dark Matter?}} \\
\vspace*{1.6cm} {\large Dan Hooper$^1$, Ignacio de la Calle Perez$^1$, Joseph
Silk$^1$, Francesc Ferrer$^1$ and Subir Sarkar$^2$} \\
\vspace*{0.5cm}
{\it $^1$Astrophysics, University of Oxford, Oxford OX1 3RH, UK\\
     $^2$Theoretical Physics, University of Oxford, Oxford OX1 3NP, UK} \\
\vspace*{0.6cm} {\tt hooper@astro.ox.ac.uk, i.calle@physics.ox.ac.uk,
silk@astro.ox.ac.uk, ferrer@thphys.ox.ac.uk, sarkar@thphys.oxford.ac.uk} \\
\vspace*{1.5cm}
\end{center}

\begin{abstract} 
\singleandabitspaced

Two ground-based experiments have recently independently detected TeV
$\gamma$-rays from the direction of the Galactic center. The
observations made by the VERITAS and CANGAROO collaborations are
unexpected, although not impossible to interpret in terms of
astrophysical sources. Here we examine in detail whether the observed
$\gamma$-rays may arise from the more exotic alternative of
annihilations of dark matter particles clustered in the center of the
Galaxy.

\end{abstract}

\end{titlepage}

\newpage
\setcounter{page}{2}
\singleandabitspaced

\section{Introduction}

Recently, the VERITAS \cite{Kosack:2004ri} and CANGAROO
\cite{Tsuchiya:2004wv} collaborations, using the Whipple 10 meter and
CANGAROO-II Atmospheric \u{C}erenkov Telescopes (ACTs), respectively,
have made significant detections of TeV $\gamma$-rays from the
Galactic center region. Although the origin of this emission is not
yet known, there are several possible, although unlikely,
astrophysical sources in the field of view. Alternatively, this may be
a signature of annihilating dark matter particles.

The Galactic center is a complex and rich region. Its most notable
inhabitant is a $2.6 \times 10^{6} M_{\odot}$ black hole, coincident
with the radio source Sgr A$^*$, which also demonstrates variable
emission at infra-red \cite{Ghez:2003hb}, soft X-ray \cite{xray} and
hard X-ray \cite{Belanger:2003se} wavelengths. Additionally, the
region may contain massive X-ray binaries emitting relativistic plasma
jets (microquasars) capable of producing high-energy $\gamma$-rays
\cite{microquasar_gamma} by either hadronic ($\pi^0$ production)
\cite{microquasar_proton} or leptonic (inverse Compton)
\cite{microquasar_lepton} processes. The region could also contain
Supernova Remnants (SNRs) which are widely believed to be the source
of Galactic cosmic rays. TeV $\gamma$-rays have indeed been observed
from several nearby supernova remnants such as the Crab
\cite{snr_crab} and Cas A \cite{snr_casa}. The responsible mechanism
remains unclear, but again could be either leptonic or hadronic (see
Ref.~\cite{Volk_snr} for a discussion of TeV $\gamma$-ray production
in SNRs). The SNR Sgr A East lies only a few parsecs away from Sgr
A$^*$, but is not a likely TeV source in itself
\cite{SgrA_East_SNR}. Interactions of its expanding shell with
molecular clouds in its environment could, in principle, produce
high-energy $\gamma$-rays \cite{Aharonian94_SNRMOLECULAR}.

Alternatively, strong winds from massive O and B type stars could lead
to hadronic interactions that may result in the emission of
high-energy $\gamma$-rays \cite{TeVOBObservations,TeVOBTheory}. Two
massive, compact, young star clusters which  contain such stars
(Arches and Quintuplet) are located roughly 10 arcminutes away from
the Galactic center. Chandra observations of the Arches cluster have
revealed non-thermal emission attributed to relativistic electrons
 accelerated in colliding wind shocks from binary systems within
the cluster or in the winds from single stars with the collective
winds from the other stars in the cluster
\cite{YusefArches1,YusefArches2}. It is argued that the existence of
non-thermal particles could result in X-ray/$\gamma$-ray emission by
inverse Compton scattering of ambient photons.

Observations by INTEGRAL \cite{INTEGRAL_SOURCES} and EGRET
\cite{dingus} have revealed $\gamma$-ray emission from the
Galactic center region, although thus far no corresponding sources
have been identified. X-ray surveys of the region have revealed a new
population of discrete sources, many of which resemble X-ray binaries
(see Ref.~\cite{Galactic_XRAY} and references therein). For more
information on the Galactic center region, see
Refs.~\cite{Galactic_XRAY,YusefZadeh00,LaRosa00}.

Despite the extensive body of evidence in favor of cold, non-baryonic,
dark matter \cite{cdm}, its identity remains elusive. Searches for
particle dark matter have been carried out using a variety of
methods. Direct searches attempt to observe the recoil energy as
Galactic dark matter particles orbiting through the Solar system
scatter elastically off nuclear targets \cite{direct}. Indirect
searches attempt to observe the products of dark matter annihilations
such as neutrinos \cite{indirectneutrino}, positrons \cite{positrons},
anti-protons \cite{antiprotons} and, in particular, high-energy
$\gamma$-rays \cite{indirectgamma,Bergstrom:1997fj,Bergstrom:2001jj}.

The remainder of this article is organized as follows. In section 2,
we summarize the observations of high-energy $\gamma$-ray emission
from the Galactic Center region. In sections 3, 4 and 5, we discuss
the spectral and spatial features of these observations and assess
whether annihilating dark matter could be responsible for
these observations. In section 6, we consider particle dark matter
candidates suggested by new physics beyond the Standard Model, in
particular supersymmetry, in the light of these recent
observations. We present our conclusions in section 7. In Refs.~\cite{Kosack:2004ri} and \cite{Tsuchiya:2004wv}, this possibility was briefly discussed. Our intention here is to explore this scenario in considerably more detail.

\section{Space and Ground Based Gamma-Ray Observations}\label{gamma_obs}

Gamma-ray observations of the Galactic Center region have been made
in several energy ranges employing a wide variety of experimental
techniques. Thus far, space-based $\gamma$-ray astronomy has been
limited to energies below 30~GeV, mainly due to the fall-off of the
photon flux at higher energies given the limited collecting area of
satellite detectors. EGRET, the Energetic Gamma-Ray Experiment
Telescope \cite{Thompson93}, launched on board the Compton Gamma-Ray
Observatory in 1991, accumulated an integrated exposure of $2 \times
10^9 \,\rm{cm}^2 \, \rm{s}$ towards the Galactic center
region. Although EGRET was sensitive in the energy range of
$\sim30~\rm{MeV}-30~\rm{GeV}$, the backgrounds are large and the
angular resolution rather poor below about 1 GeV. EGRET detected a
strong source of GeV $\gamma$-rays in the Galactic center region,
although this source appears to be about 10 arcminutes away from the
dynamical center of the Galaxy (Sgr A$^*$) \cite{dingus}. Only an
upper limit could be placed on the $\gamma$-ray flux from Sgr A$^*$
itself.
EGRET's successor, GLAST (the Gamma-ray Large Area Space Telescope),
is planned for launch in 2006. GLAST will provide a sensitivity of
about 50 times that of EGRET (above 100~MeV), a factor of $\sim5$ increase in
effective area, better angular resolution (less than 0.1$^\circ$ above
10 GeV), and will be sensitive to far higher energies ($\sim300$ GeV)
with an energy resolution better than 15\% even at the high-energy end
\cite{Gehrels:ri}.

In addition to satellite-based experiments, ground-based ACTs provide
measurements from $\sim200$~ GeV up to $\sim10$ TeV. Typical
characteristics for current ACTs are effective areas of $\sim
10^5~\rm{m}^2$ (dependent on the zenith angle), peak response energies
down to a few hundred GeV (dependent on both the zenith angle of the
source and its energy spectrum), fields-of-view of a few degrees,
angular resolutions of $0.1^{\circ}$ to $0.2^{\circ}$ and
sensitivities down to a few hundredths of the Crab Nebula's flux. The
detection technique relies on using the Earth's atmosphere as a
calorimeter. One of the components of extensive air showers triggered
by $\gamma$-rays and cosmic-rays in the atmosphere is \u{C}erenkov
radiation produced by the charged component of the shower. Differences
in the longitudinal and transverse development between $\gamma$-ray
and cosmic-ray induced showers are reflected in the \u{C}erenkov
images recorded by a high resolution camera located on the focal plane
of a reflector (see, for example, Ref.~\cite{Hillas96}). These
differences enable rejection of over 99\% of the `unwanted' cosmic-ray
events (see, for example, \cite{Reynolds:av}). In the remaining
images, the energy and arrival direction of the primary $\gamma$-ray
can be deduced.

Detections of the Galactic center at very high energies have been
reported by two ACTs in the past few weeks. The VERITAS
collaboration has made observations with the Whipple 10\,m telescope
\cite{whipple} on Mt. Hopkins, Arizona. Due to its northern location
(31$^\circ$ 57.6' N), these observations had to be made at high zenith
angle, resulting in a rather high energy threshold of
$\sim2.8\,$TeV. Between 1995 and 2003, 26 hours of data were
accumulated from this direction, resulting in a 3.7$~\sigma$ signal
with an integral flux of \cite{Kosack:2004ri}
\begin{equation}
 F_\gamma (> 2.8\ \rm{TeV}) = 1.6 \pm 0.5\ (\rm{stat}) \pm 0.3\
                           (\rm{syst}) \times 10^{-8}\ \rm{photons\
                           m}^{-2}\ \rm{s}^{-1}.
\label{whippleflux}
\end{equation}
This corresponds to $\sim40\%$ of the Crab Nebula flux (the `standard
candle' in TeV $\gamma$-ray astronomy) above the same energy.

For ACTs located in the southern hemisphere, the energy threshold in
the direction of the Galactic center is an order of magnitude lower
than that of Whipple. Observations taken during 2001 and 2002 with
CANGAROO-II \cite{Kawachi01} have yielded a $\sim 10\sigma$ detection
of the Galactic Center region above $\sim 250$ GeV
\cite{Tsuchiya:2004wv}, with an integrated flux of
\begin{equation}
 F_\gamma (> 250\ \rm{GeV}) \simeq 2 \times 10^{-6}\, \rm{photons\
                                   m}^{-2}\ \rm{s}^{-1}.
\label{cangarooflux}
\end{equation}
These measurements indicate a very soft spectrum ($\propto
E^{-4.6\pm0.5}$) and a flux at 1~TeV corresponding to 10\% of the Crab
Nebula flux.

\section{Annihilating Dark Matter: Spectral Characteristics}

\begin{figure}[t]
\centering\leavevmode \mbox{
\includegraphics[width=3.2in]{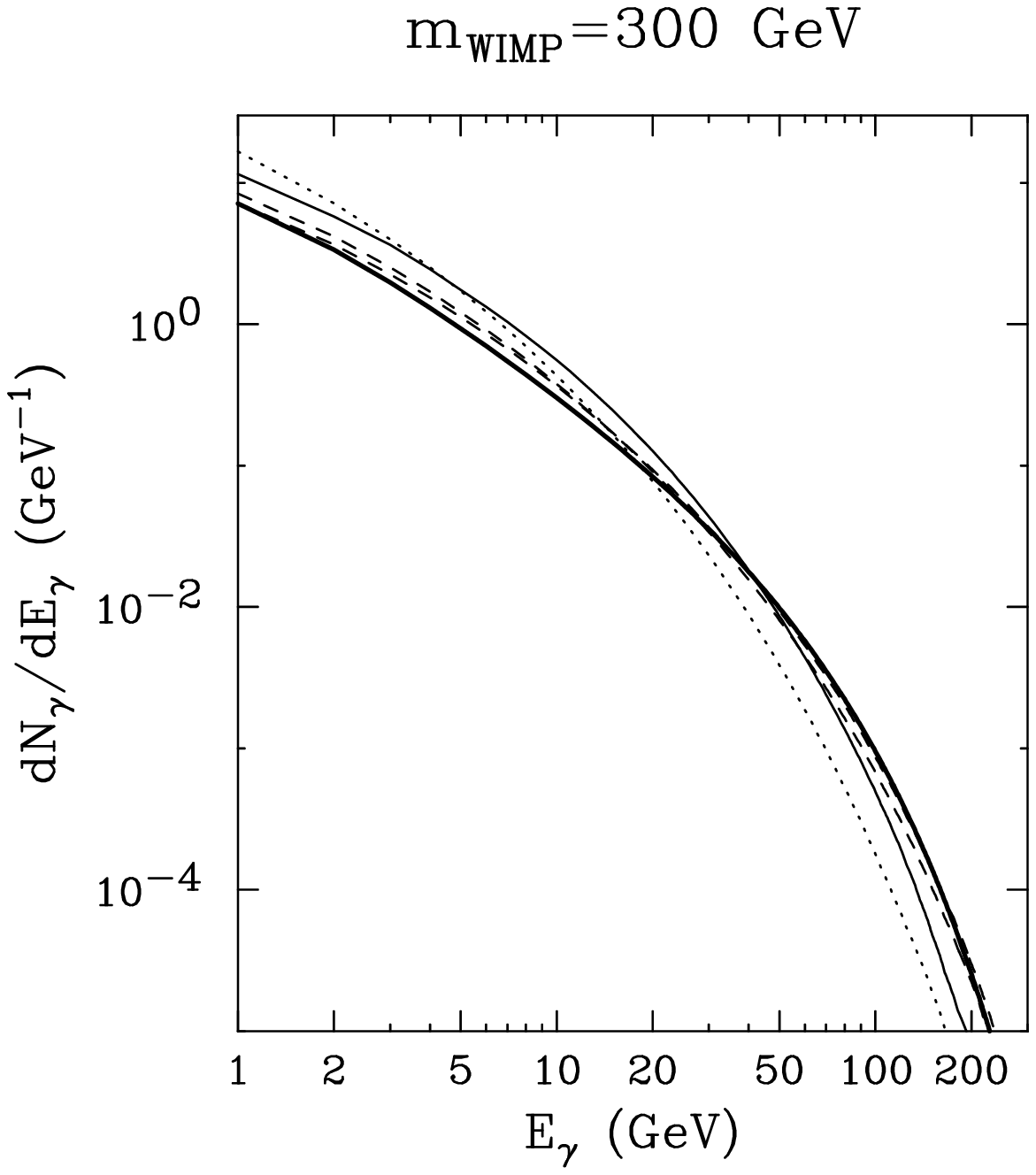} \hfill
\includegraphics[width=3.2in]{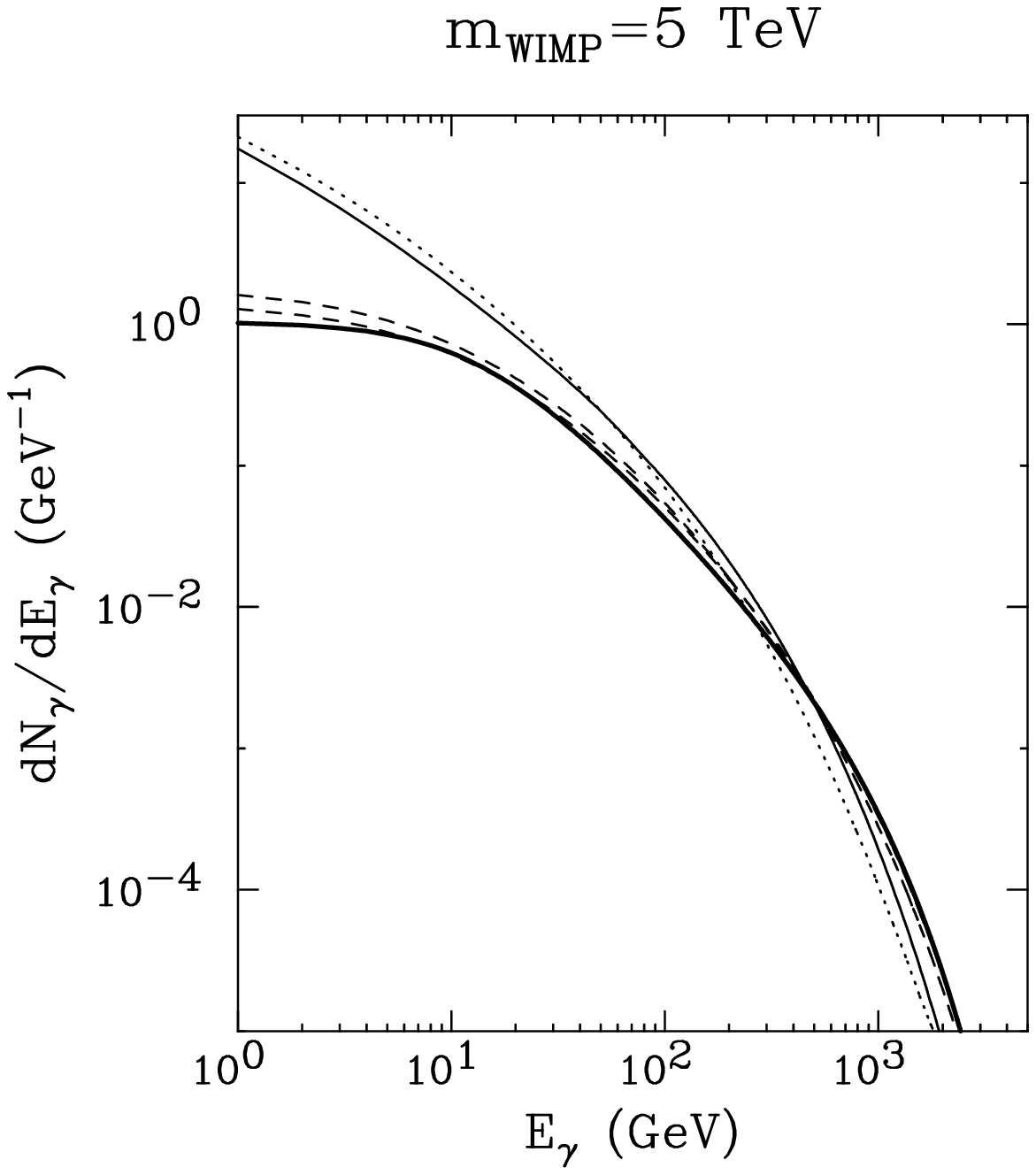}}
\caption{The spectrum of $\gamma$-rays for dark matter annihilation to
selected modes --- $b\bar{b}$ (thin full line), $t\bar{t}$ (dotted
line), $W^+W^-, ZZ$ (dashed line). The parameterization of
Eq.~\ref{para} is also shown (thick full line).}
\label{spec}
\end{figure}

The annihilations of dark matter particles can produce $\gamma$-rays
in several ways. First, a continuum of $\gamma$-rays results from the
hadronization and decay of $\pi^0$'s generated in the cascading of
annihilation products. Second, monoenergetic $\gamma$-ray lines are
produced as dark matter particles annihilate via the modes $X X
\rightarrow \gamma \gamma$ and $X X \rightarrow \gamma Z$. However the
Feynmann diagrams for line-producing processes typically involve loops
and thus yield much smaller fluxes than continuum emission.
The spectrum of $\gamma$-rays from continuum emission depends on which
annihilation modes dominate. Annihilations to light quark pairs result
in a fairly hard spectrum, while the spectrum is somewhat softer for
heavy quarks ({\it i.e.} $t \bar{t},\, b \bar{b}$). Annihilations to
gauge bosons demonstrates behavior in between these cases. In the
energy range above $\sim10^{-2}$ times the dark matter particle mass,
$M_X$, these variations are mild. At lower energies, the $\gamma$-ray
spectrum from gauge boson modes flattens (see Fig.~\ref{spec}).

For dark matter annihilations to gauge boson pairs, the resulting
$\gamma$-ray spectrum can be parameterized as (see
Refs.~\cite{Bergstrom:1997fj,Bergstrom:2001jj})
\begin{equation}
\label{para}
\frac{{\rm d}N_\gamma}{{\rm d}E_\gamma} \simeq \frac{0.73}{M_X}
 \frac{{\rm e}^{-7.76 E_\gamma/M_X}}{(E_\gamma/M_X)^{1.5} + 0.00014}.
\end{equation} 
In Fig.~\ref{spec}, this parameterization is compared to the spectrum
obtained using the PYTHIA fragmentation Monte Carlo \cite{pythia} as
implemented in the DarkSusy programme
\cite{darksusy}.\footnote{Similar results are obtained using the
HERWIG fragmentation Monte Carlo \cite{Birkel:1998nx}, as well as by
direct evolution of the fragmentation functions measured at LEP using
the DGLAP equations \cite{Sarkar:2001se}.}  The parameterization of
Eq.~\ref{para} is reasonably accurate for particles with masses in the
range we will be concerned with here. If annihilations into heavy
quarks are important, the spectrum will be modified --- annihilations
to $b \bar{b}$ and $t \bar{t}$ will produce fewer $\gamma$-rays in the
energy range $0.1 M_{X} < E_{\gamma} < M_{X}$, but more $\gamma$-rays
at lower energies.

Alternatively, the processes $X X \rightarrow \gamma
\gamma$ and $X X \rightarrow \gamma Z$ produce $\gamma$-rays with energies $M_X$ and $M_X (1- M_Z^2/4 M_X^2)$, respectively. The flux of these lines is determined by the appropriate cross-sections for each
process, which vary from model to model. For the range of possible
cross-sections for line emission in supersymmetric models, see
Ref.~\cite{Bergstrom:1997fj}. The flux of these lines is typically
much smaller than the corresponding continuum flux.

\begin{figure}[t]
\centering\leavevmode \mbox{
\includegraphics[width=3.5in]{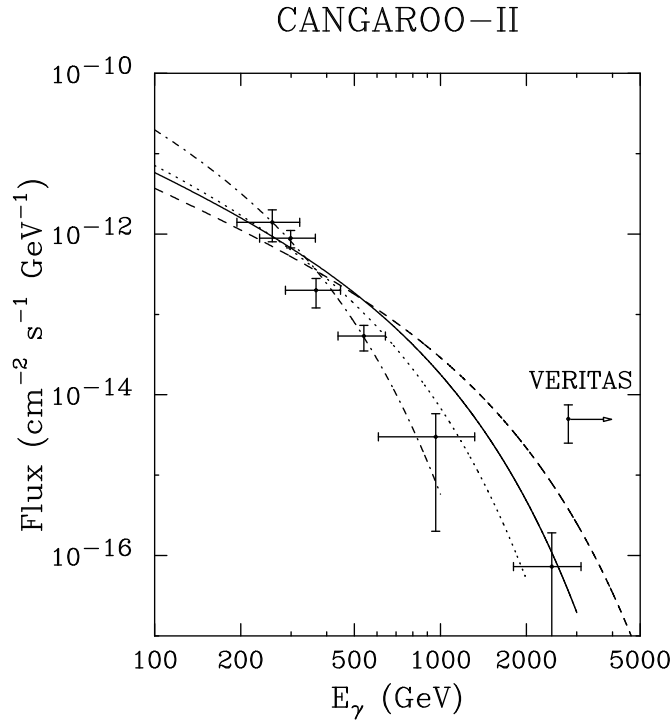} }
\caption{Data from the CANGAROO-II experiment compared with the
spectrum predicted for dark matter annihilations to gauge bosons (see
Eq.~\ref{para}). The dot-dashed, dotted, solid and dashed lines are
for 1, 2, 3 and 5 TeV particles. Normalization was considered a free
parameter. Note that the highest energy bin shown (near 2.5 TeV) is
less than 1 $\sigma$ in excess of a null result, so should be taken
only as an upper limit. Also shown is the flux measured by the VERITAS
collaboration, inferred from the integral flux assuming the spectrum
of a 5 TeV mass annihilating particle. Note the very different results
of the two experiments.}
\label{specdata}
\end{figure}

The CANGAROO-II experiment has published their spectrum in six energy
bins. These results are shown in Fig.~\ref{specdata} in comparison to
the spectrum predicted by Eq.~\ref{para}. The data from CANGAROO-II
appears to fit the spectrum reasonably well for a 1--3 TeV dark matter
particle. For a heavier particle, the spectrum measured by CANGAROO-II
is much softer than would be predicted for annihilating dark matter.
This is also in conflict with the Whipple experiment which finds a
substantial flux above 2.8 TeV.

\begin{figure}[t]
\centering\leavevmode \includegraphics[width=3.5in]{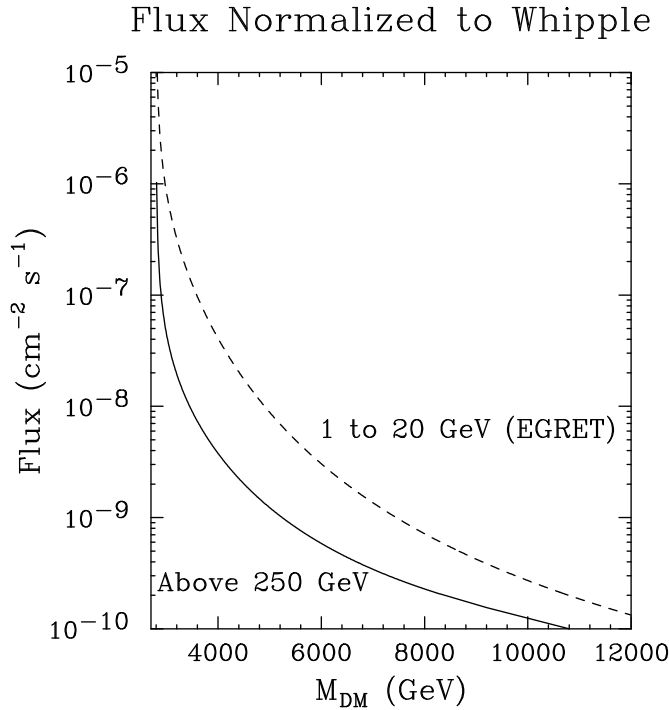}
\caption{The integrated flux predicted for CANGAROO-II, HESS (with 250 GeV thresholds) and EGRET
if annihilating dark matter is the source of the $\gamma$-rays
observed by Whipple. Annihilations primarily to gauge bosons are
assumed (using the parameterization of Eq.~\ref{para}). With
CANGAROO-II's integrated flux of $\sim 2 \times 10^{-10} \,
\rm{cm}^{-2}\,\rm{s}^{-1}$ above 250 GeV, only a very heavy dark matter
particle ($\sim 10$ TeV) is consistent with the CANGAROO-II and
Whipple results, assuming continuum emission dominates. If line
emission is significant, the mass may be somewhat smaller. If the
particle mass is less than 3.5--4 TeV, the continuum emission from
annihilations into gauge bosons exceeds the limit placed by EGRET
\cite{dingus}, assuming a negligible line component. If dark matter
annihilates mostly to another mode, such as heavy quark pairs, the
EGRET limit may be violated for WIMPs as heavy as $\sim5$ TeV.}
\label{compare}
\end{figure}

Given the integrated flux above 2.8 TeV recorded by Whipple
(Eq.~\ref{whippleflux}), the corresponding flux expected for EGRET,
CANGAROO-II or HESS can be estimated for a given dark matter particle
mass using the parameterization of Eq.~\ref{para}. As seen in
Fig.~\ref{compare}, the integrated flux observed by CANGAROO-II
(Eq.~\ref{cangarooflux}) is consistent with the Whipple measurement
only for a very heavy particle ($\sim 10$~TeV) which produces
$\gamma$-rays primarily by continuum emission. This appears to be in
contradiction with the results of Fig.~\ref{specdata}. A somewhat lighter
particle may be accomodated if the observed $\gamma$-rays have a
significant component of line emission, however.

If dark matter annihilation indeed produces the TeV $\gamma$-rays
observed by ACTs, then a lower energy component is expected to which
EGRET (or in the future GLAST) is, in principle, sensitive. However
EGRET has placed only an upper limit on the $\gamma$-ray flux from the
Galactic center above 1 GeV
\cite{dingus}. This corresponds to an annihilation flux upper limit of
$\sim 10^{-8}\,\rm{cm}^{-2}\, \rm{s}^{-1}$ for a 100 GeV dark matter
particle, and $\sim 10^{-7}\,\rm{cm}^{-2}\, \rm{s}^{-1}$ for a multi-TeV
particle. Thus we see from Fig.~\ref{compare} that a spectrum
normalized to the Whipple observations will {\em violate} the EGRET
bound if the particle mass is below about 3.5--4 TeV. If annihilations
to modes other than gauge bosons dominate, this bound excludes masses
up to about 5 TeV. On the other hand, if line emission is substantial,
the corresponding flux in the range of EGRET's sensitivity would be
reduced.

\section{The Annihilation Rate}

The flux of $\gamma$-rays observed from the Galactic center, if
interpreted as due to dark matter annihilation, can be used to
constrain the annihilation cross-section of a dark matter candidate
particle and the dark matter halo profile characteristics. The
$\gamma$-ray flux from dark matter annihilations near the Galactic
center is given by
\begin{equation}
\Phi_\gamma (\psi, E_\gamma) = \langle\sigma v\rangle
                               \frac{\rm{d}N_\gamma}{\rm{d} E_\gamma}
                               \frac{1}{4\pi M_X^2}\int_{\rm los}
                               \rm{d}l(\psi)\ \rho^2(r).
\end{equation}
Here, $\psi$ is the angle between the line-of-sight (los) and the
Galactic center, $\langle\sigma v\rangle$ is the dark matter
annihilation cross-section averaged over its velocity distribution,
and $\rho(r)$ is the dark matter density at distance, $r$, from the
Galactic center. This expression can be conveniently separated into
two factors --- the first specifying the particle physics model (mass,
cross-section and fragmentation spectrum), and the second describing
the dark matter distribution. Normalizing to the distance to the
Galactic center and the local halo dark matter density, the latter
factor can be written as:
\begin{equation}
 J (\psi) = \frac{1}{8.5\,\rm{kpc}}
            \left(\frac{1}{0.3\,\rm{GeV/cm}^3}\right)^2 \int_{\rm los}
            \rm{d}l(\psi)\ \rho^2(l),
\end{equation}
so we can calculate this factor for any assumed dark matter distribution near Galactic center. Then, defining $\overline{J(\Delta \Omega)}$ as the average of
$J(\psi)$ over the solid angle $\Delta \Omega$ (centered on $\psi=0$),
we can write
\begin{equation}
 \Phi_\gamma (\psi, E_\gamma) \simeq 5.6 \times 10^{-12}\
 \rm{cm}^{-2}\rm{s}^{-1} \frac{\rm{d}N_\gamma}{\rm{d}E_\gamma}
 \left(\frac{\langle\sigma
 v\rangle}{3\times10^{-26}\rm{cm}^3\rm{s}^{-1}}\right)
 \left(\frac{M_X}{1\,\rm{TeV}}\right)^{-2} \overline{J (\Delta\Omega)}
 \Delta\Omega.
\end{equation}
For ACTs, which typically have angular resolutions of order
$0.2^\circ$, we will consider a solid angle of $\Delta\Omega \sim
5\times10^{-5}$ sr. The value of $\overline{J(\Delta\Omega)}$ can vary
a great deal depending on the assumed dark matter distribution,
e.g. for an NFW halo profile \cite{nfw} $\overline{J(5 \times
10^{-5}\, \rm{sr})} \simeq 5.6 \times 10^3$, while a Moore {\it et
al.} halo profile \cite{moore} yields a considerably larger value,
$\overline{J(5 \times 10^{-5}\, \rm{sr})} \simeq 1.9 \times 10^6$.

It is possible, however, that these highly cusped profiles deduced
from N-body simulations \cite{nbody}, do not accurately represent the
distribution of dark matter in our halo. In particular, current N-body
simulations cannot resolve halo profiles on scales smaller than
roughly 1 kpc, and must rely on extrapolations in the innermost
regions of the Galactic center \cite{Binney:2003sn}. Also, these
simulations model halos without baryons, so may not be valid in the
inner region of the Galaxy which is baryon dominated. If the baryons
in the inner halo had been significantly heated in the past, they may
have expanded outward, gravitationally pulling the dark matter and
thus reducing its density in the innermost regions. Conversely, as
baryonic matter loses energy through radiative processes, it will fall
deeper into the Galaxy's gravitational well, pulling dark matter along
with it; halo models which include this `adiabatic compression'
effect predict substantially higher dark matter densities near the
Galactic center \cite{klypin}.  Alternatively, adiabatic accretion of
dark matter onto the Super-Massive Black Hole (SMBH) at the center of our
Galaxy occuring as a consequence of adiabatic growth of the SMBH
may have produced a density `spike' in the halo profile
\cite{spike}. If this is the case, a very bright $\gamma$-ray source
could be produced from dark matter annihilations. Such a spike
could have been destroyed in a series of hierarchical mergers
\cite{merrit} although such mergers are unlikely to have occurred in
the recent history of the Milky Way. The spike would most
likely have been modified in the earliest stages of SMBH growth, however,
when some mergers should have occurred.
A more likely continual source of softening
of the spike is the effect of
stellar encounters \cite{merritta}. However the annihilation flux is
still enhanced in this case relative to the cusp in the absence of the SMBH.
Moreover such a spike may provide the source of
relativistic electrons needed to account for the low frequency radio
emission from Sgr A$^*$ \cite{bertone}. There is no merit in a recent 
claim
that the  density spike is inconsistent with radio observations of the
Galactic Center, since this study only considered the case of an
initial NFW profile; the processes described above will inevitably
soften the profile \cite{abo}. It is even possible that other 
massive black holes are present in the Galactic center region, being
failed mergers that are relics of the ``final parsec'' problem \cite{milos},
as predicted by hierarchical merging scenarios that are normalised to
the
observed SMBH mass-spheroid velocity dispersion relation 
\cite{madau, islam}. These objects, which cumulatively contain as much mass as
the
central SMBH, would most likely have retained their CDM spikes
generated if they  formed adiabatically in the cores of
pregalactic dwarf Galaxies, and hence are potential
$\gamma$-ray sources.

There are other observations which can be used to constrain the dark
matter distribution. In particular, by studying microlensing events in
the direction of the Galactic Bulge, the quantity of dark matter
within the Solar circle can be constrained. Binney and Evans \cite{be}
have argued that the observed number of microlensing events can be
used to exclude cuspy halo profiles with inner power-law indices
greater than about 0.3; for comparison, NFW and Moore {\it et al.}
profiles have indices of 1.0 and 1.5, respectively. However, Klypin
{\em et al.}  claim to find reasonable agreement between observational
data and cuspy halo profiles \cite{kzs} and argue furthermore that
there is no conflict with microlensing data if adibatic compression is
included \cite{klypin}.  These authors also study  
the effects of angular momentum transfer from a fast rotating central
bar which can help  diminish the spike.
We note that their different conclusion
is in large part due to the adoption  of a
significantly  lower microlensing optical depth towards the GC
than considered by Binney and Evans.
Given this wide range of opinions regarding
the halo dark matter profile, it may be prudent not to exclude any of
these models from our discussion.

\begin{figure}[t]
\centering\leavevmode \mbox{
\includegraphics[width=3.2in]{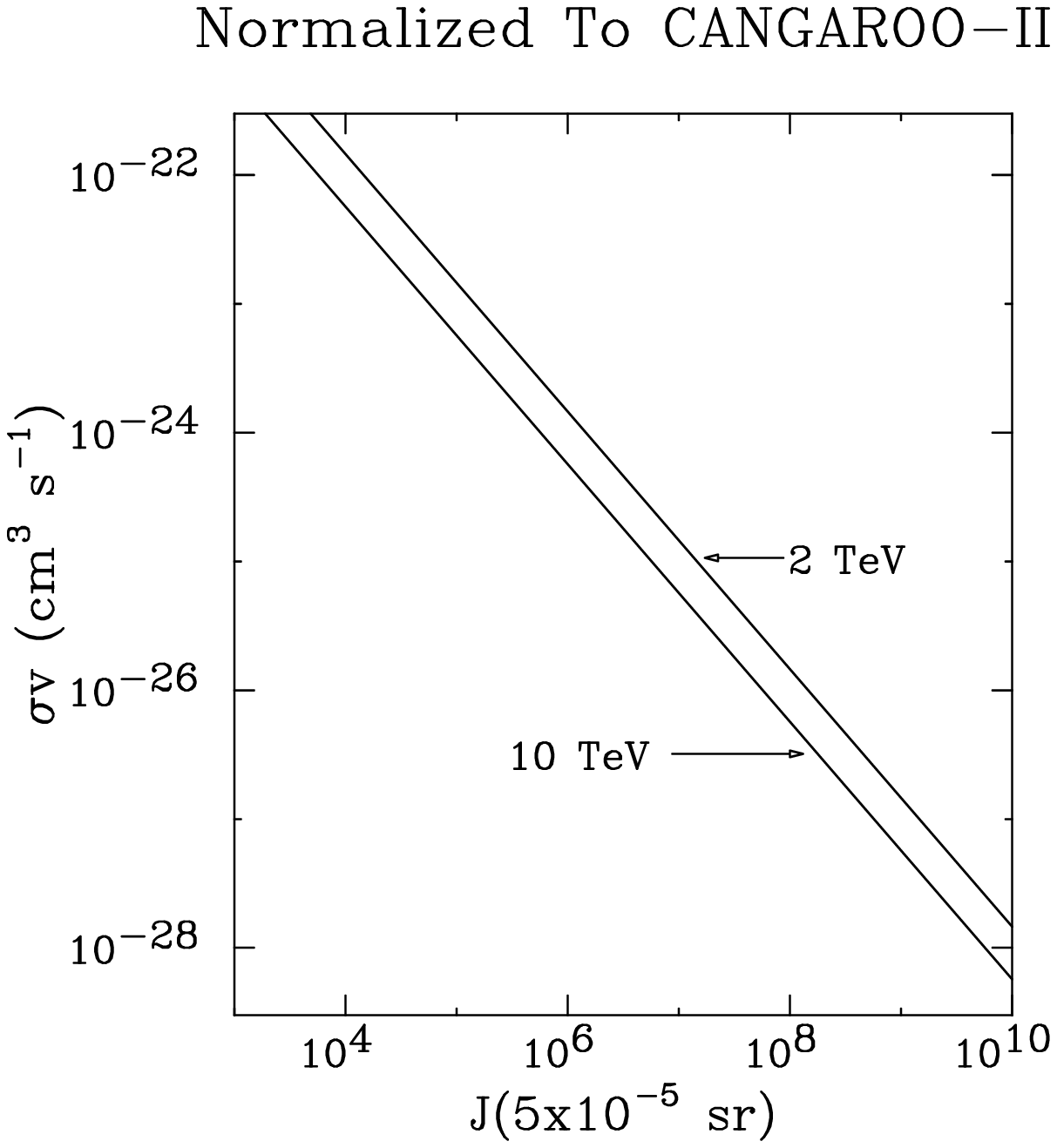} \hfill
\includegraphics[width=3.2in]{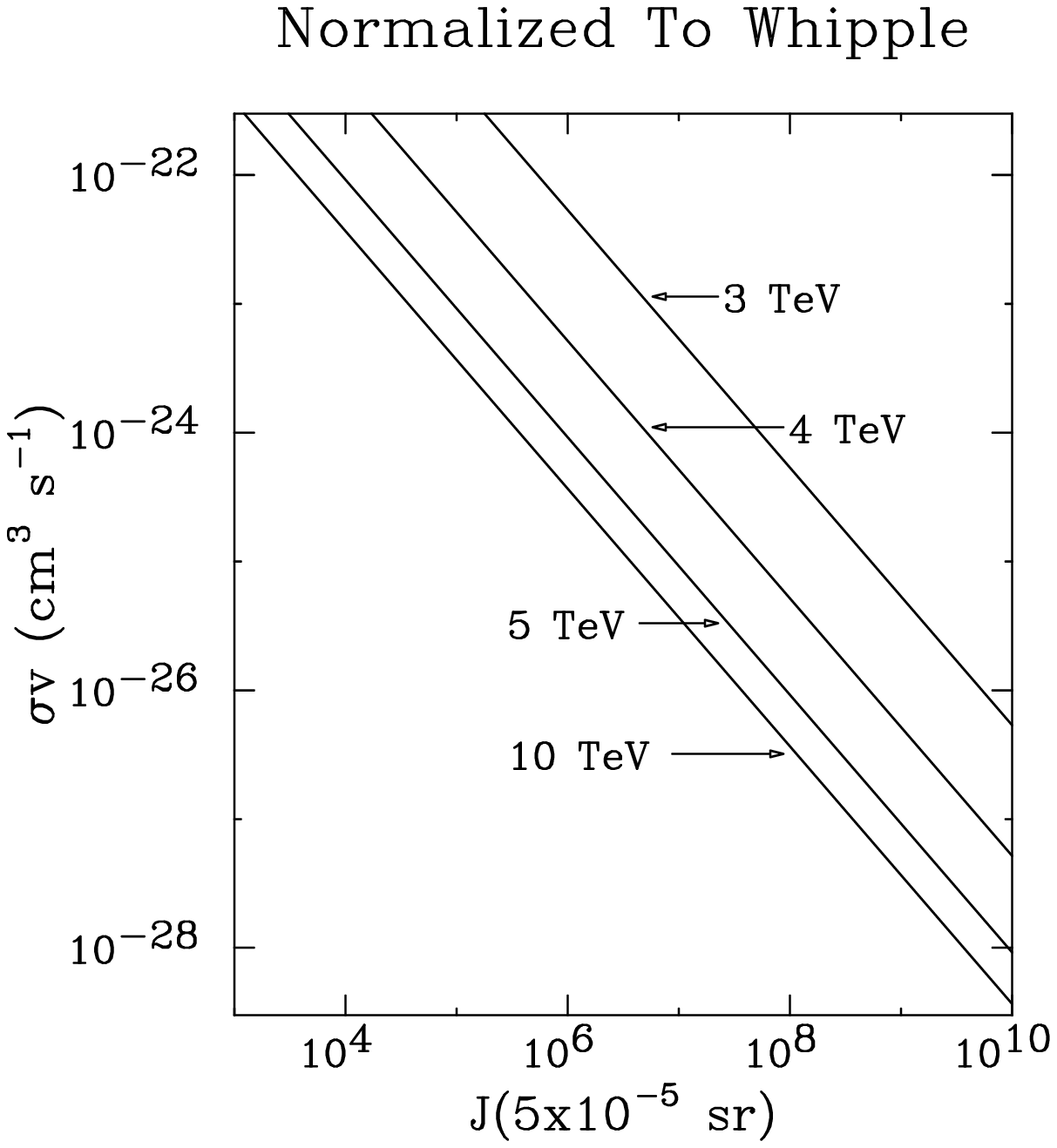}}
\caption{The annihilation cross-section and halo profile needed to
provide the $\gamma$-ray flux observed by CANGAROO-II (left) and
Whipple (right). Contours are shown for several dark matter particle
masses.}
\label{rate}
\end{figure}

We now turn our attention to the parameters set by the particle
physics of the dark matter candidate, in particular the annihilation
cross-section which also determines its relic thermal abundance. In
the simplest situation, where the annihilation cross-section is {\em
independent} of velocity, the relic abundance is approximately given
by
\begin{equation}
 \Omega_X h^2 \sim \left(\frac{\langle\sigma
v\rangle}{3\times10^{-27}\,\rm{cm}^{3}\rm{s}^{-1}}\right)^{-1}.
\end{equation}
Of course, the annihilation cross-section is generally not
velocity-independent. If it is larger at high velocities, the
cross-section at low velocities will need to be smaller to yield the same
relic abundance. Similarly, if resonances or co-annihilations
significantly reduce the relic abundance, a smaller annihilation cross-section at low velocities will be required to match the measured
density of dark matter. Although there are certainly exceptions to
this estimate, it is reasonable to consider $3 \times 10^{-26} \,
\rm{cm}^3 \, \rm{s}^{-1}$ 
(or $\sim$ 10 pb) as an upper limit for the (low velocity)
annihilation cross-section for a thermal relic which makes up the cold
dark matter. Of course, this limit could be exceeded if the relic
abundance did not result from an initial state of thermal
equlibrium. A non-thermally produced dark matter particle could have a
considerably larger annihilation cross-section.

In figure~\ref{rate}, we show the annihilation cross-sections and
values of $\overline{J(5 \times 10^{-5}\, \rm{sr})}$ needed to
accomodate the CANGAROO-II (left) and Whipple (right)
observations. For an NFW halo profile ($\overline{J(5 \times 10^{-5}\,
\rm{sr})}\sim 10^4$), very large cross-sections of $\sim 10^{-23} \,
\rm{cm}^3 \, \rm{s}^{-1}$ would be required to match the fluxes
detected by Whipple or CANGAROO-II. A thermal relic with such a large
cross-section would not yield a significant relic abundance. On the
other hand, if we consider a more centrally concentrated halo
distribution, such as a Moore {\it et al.} profile with adiabatic
compression, $\overline{J(5 \times 10^{-5}\, \rm{sr})}$ could be as
large as $\sim10^8$. In this case, cross-sections on the order  of
$\sim 3 \times 10^{-26} \, \rm{cm}^3 \, \rm{s}^{-1}$ could suffice to
match the observed fluxes. Spiked density profiles could readily
accommodate the observed flux.  Indeed, one could even
dispense with the need for the spike to surround the central SMBH, 
provided that another pregalactic  VMBH  which had retained its spike 
would be within a few parsecs of
the GC.
\section{The Dark Matter Distribution}

\begin{figure}[t]
\centering\leavevmode \includegraphics[width=3.7in]{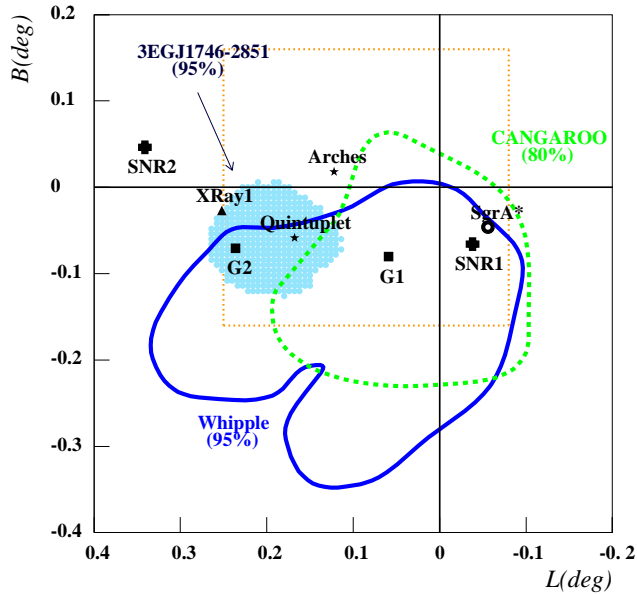}
\caption{A skymap of the Galactic center region. The solid and dashed
contours correspond to the regions observed by Whipple and
CANGAROO-II, respectively. In these regions the observed significance
is greater than 95\% for Whipple and 80\% for CANGAROO-II. The 95\%
confidence region for the off-center source observed by EGRET (3EG
J1746-2851) is shown as a shaded region. Also shown are a number of
selected objects known to be present in the region including Sgr A$^*$
(the dynamical center of the Galaxy and location of the supermassive
black hole), two supernova remnants (SNR1 and SNR2, corresponding to
Sgr A East and SNR 000.3+00.0, respectively), the Arches and
Quintuplet star clusters, the low mass X-ray binary 1E 1743.1-2843 and
two $\gamma$-ray sources observed by INTEGRAL (G1 and G2). The boxed area is the region shown in Fig.~\ref{skymapxray}}
\label{skymap}
\end{figure}

\begin{figure}[t]
\centering\leavevmode \includegraphics[width=4.7in]{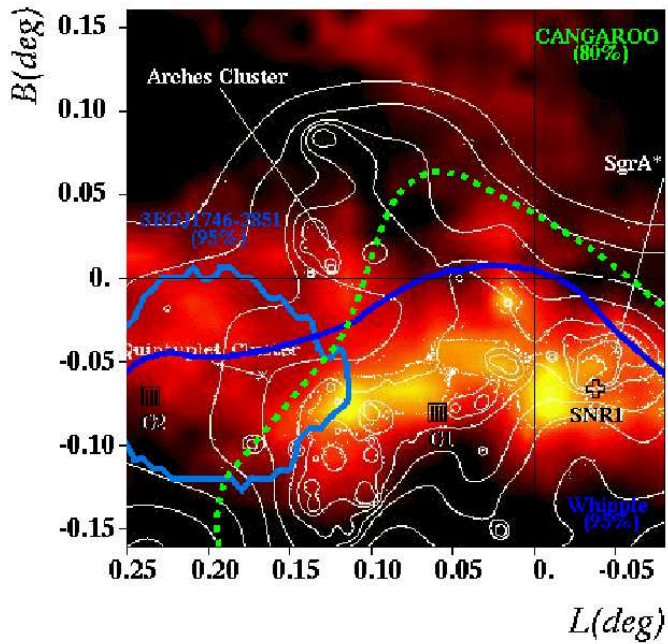}
\caption{A skymap of the Galactic center region overlaid with the
results of the Chandra X-ray telescope at 6.4 keV (shading) and radio
observations (thin contours) \cite{jackson}. The objects, thick
contours and regions shown are the same as in Fig.~\ref{skymap}. The 95\% confidence region observed by EGRET is shown as a solid contour in the lower left quadrant of the figure. The
X-ray/radio map is provided courtesy of Q.~D.~Wang.}
\label{skymapxray}
\end{figure}

The observations of the Galactic center made by Whipple and
CANGAROO-II are consistent with emission from a point source. These
experiments have angular resolutions of order 10 arcminutes,
however, so the source may be extended roughly up to this angular scale. Each experiment's result is also
consistent, to within their angular resolutions, with emission from
the Galaxy's dynamical center. A skymap showing the regions
corresponding to the CANGAROO-II and Whipple detections is shown in
Fig.~\ref{skymap}. In Fig.~\ref{skymapxray}, a similar map is shown
including X-ray and radio observations of the region.

The angular distribution of $\gamma$-rays from dark matter
annihilations can be calculated for a given halo profile. At one
extreme, profiles with a density spike predict an angular distribution which is essentially a point source with a negligible
width. At the other extreme, a profile with a flat core in the inner
few kiloparsecs will imply a rather extended distribution which cannot be reconciled with the recent observations
by ACTs.

\begin{figure}[t]
\centering\leavevmode \includegraphics[width=3.5in]{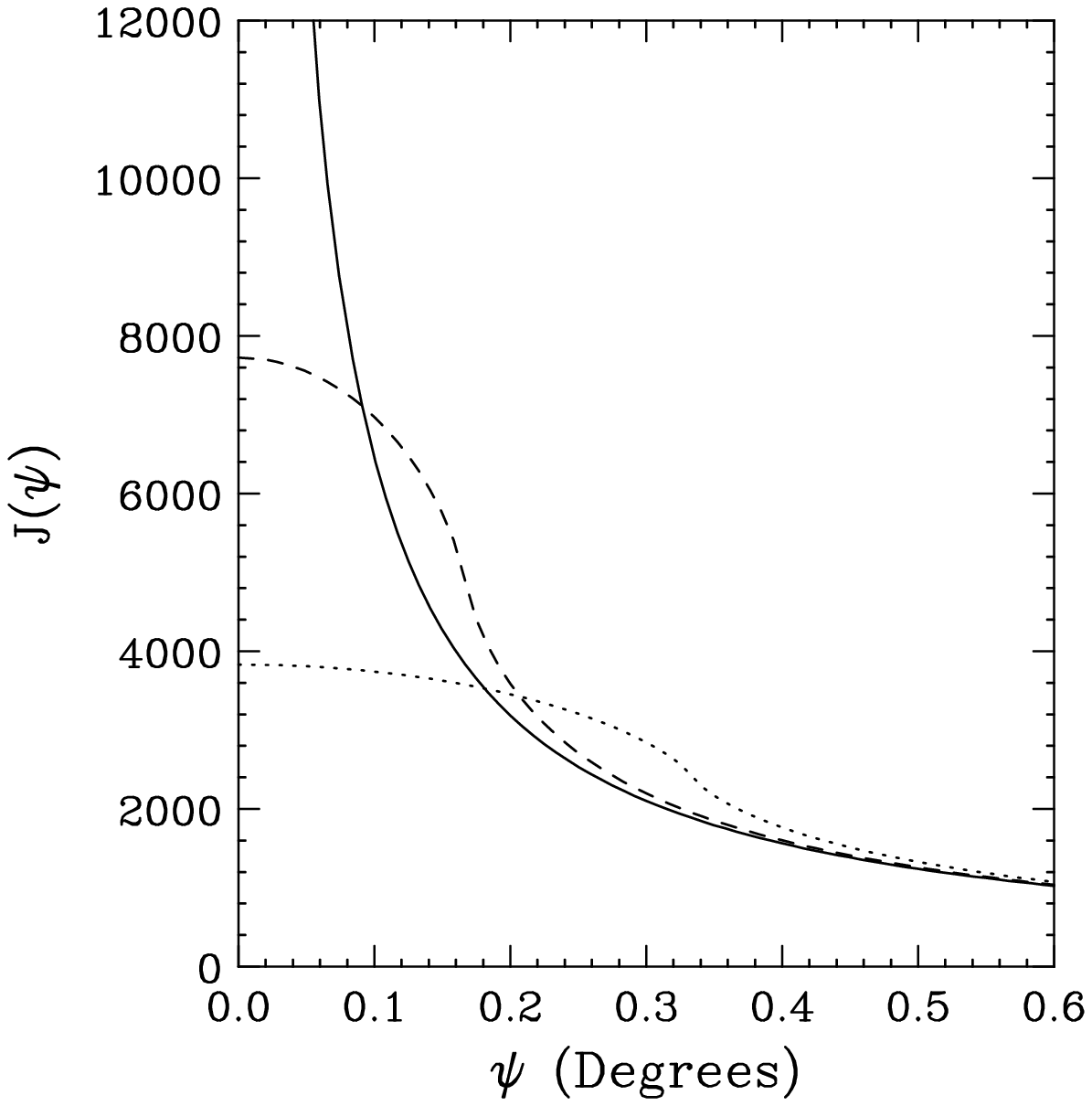}
\caption{The angular distribution of $\gamma$-rays predicted from dark
matter annihilations assuming an NFW halo profile, for an experiment
with sub-arcminute (solid line), 10 arcminute (dashed line) and 20
arcminute (dotted line) angular resolution. Current ACT observations
of the Galactic center cannot differentiate a cuspy halo profile from
a point source.}
\label{angular}
\end{figure}

Cusped halo profiles fall between these extremes. In
Fig.~\ref{angular}, we show the angular distribution of $\gamma$-rays
predicted using an NFW halo profile for experiments with
sub-arcminute, 10 arcminute and 20 arcminute resolution. It is clear
that with present data, it is impossible to differentiate between an
NFW halo profile and a point source. The same conclusion is reached if
other cusped distributions or adibatically compressed profiles are
considered.

There is, of course, the possibility that the $\gamma$-rays observed
by Whipple and CANGAROO-II are {\em not} from dark matter
annihilations, but rather originate from some astrophysical source.
In particular, there exists a source of 100 MeV to 15 GeV
$\gamma$-rays which was observed by EGRET approximately 0.2$^{\circ}$
away from the Galactic center (see figure~\ref{skymap}). Given their
limited angular resolution, this location is {\em consistent} with the
observations made by Whipple and CANGAROO-II. Future experiments with
improved angular resolution will be needed to determine if the TeV
emission from this region originates from the Galaxy's dynamical
center, or from an offset direction. In particular, the HESS
experiment has begun operating with four telescopes which should
improve the angular resolution by a factor of 2 over a single
telescope. GLAST will also have substantially better angular
resolution than its predecessor EGRET.

\section{Particle Dark Matter Candidates}

For annihilating dark matter to accommodate the recent observations of
ACTs, Whipple in particular, the dark matter candidate must be a few
TeV or heavier. In models of softly broken supersymmetry, the lightest
supersymmetric particle is not usually expected to be so heavy, being
more natural near the electroweak scale. However, TeV-scale masses are not necessarily fine tuned, {\it e.g.}, in
the focus point region of the constrained minimal supersymmetric
standard model (sometimes called mSUGRA), neutralinos have such masses \cite{focusnatural}. In such models, the lightest neutralino
is typically a higgsino or a mixed gaugino-higgsino, and thus
annihilates mostly to gauge boson pairs (rather than heavy quarks).

Outside of the focus point region, multi-TeV neutralinos can provide
the observed dark matter relic density only in special scenarios. For
example, if the CP-odd Higgs boson is very nearly twice the mass of
the lightest neutralino, annihilations can occur at resonance, again
allowing for TeV scale dark matter \cite{resonances}. This situation
can arise in the large $\tan \beta$ region of radiative gauge symmetry
breaking models, such as mSUGRA. Alternatively, if the lightest
neutralino is nearly degenerate with another supersymmetric particle
(such as a stop or stau), coannihilations can deplete the thermal
relic density, thus allowing for heavy neutralino dark matter
\cite{coannihilations}. It should be said that despite these
possibilities, supersymmetry is typically expected to appear well
below the TeV-scale.

The $\gamma$-ray signatures for supersymmetric dark matter have been
studied extensively in the literature (see, for example,
Refs.~\cite{indirectgamma,Bergstrom:1997fj,Bergstrom:2001jj,dingus,loop}).
If a multi-TeV neutralino is indeed responsible for the observed
$\gamma$-ray emission, supersymmetry will be very difficult to study
in accelerator experiments, such as the Large Hadron Collider (LHC),
making astro-particle experiments a more viable probe.

Thus far, we have only discussed dark matter which is a thermal
relic. In neutralinos are produced in non-thermal processes well after
the freeze-out epoch, the dark matter candidate may have a much larger
annihilation cross-section than for the case of a thermal relic. For
example, in anomaly-mediated supersymmetry breaking, the lightest
neutralino is Wino-like with annihilation cross-sections larger
than in most other supersymmetry scenarios. Such a dark matter candidate may require non-thermal
mechanisms to generate the observed dark matter density \cite{amsb}.

In addition to supersymmetry, other extensions of the Standard Model
can also provide a viable dark matter candidate. These include
Kaluza-Klein dark matter in models with universal extra dimension
\cite{gammakk}, scalar dark matter in 'theory space little Higgs'
models \cite{lhdm} and models with very heavy neutrinos
\cite{heavynu}. Although there are numerous other examples of potentially interesting dark matter candidates discussed in the literature, it is
beyond the scope of this paper to discuss them here.

\section{Summary and Conclusions}

In this article, we have discussed the possibility that annihilating
dark matter in the Galactic center has produced the flux of
$\gamma$-rays observed by the CANGAROO and VERITAS
collaborations. Although it is possible that these $\gamma$-rays are
the result of astrophysical processes, we summarize here the
characteristics required of a dark matter particle, if its
annihilations are responsible for the observed $\gamma$-ray emission.

\subsection{The Spectrum}

It is difficult to reconcile the spectra observed by the CANGAROO-II
and Whipple experiments. The spectrum measured by CANGAROO-II is
consistent with an annihilating particle of mass in the range
of 1--3 TeV. On the other hand, Whipple has observed a substantial
flux above its rather high threshold of 2.8 TeV, requiring a much
heavier dark matter particle. Future observations will be needed to
conclusively determine the spectrum of $\gamma$-rays from the Galactic
center in the GeV-TeV range.

\subsection{The Halo Profile and Annihilation Cross Section}

For annihilating dark matter to produce the flux measured by either
the CANGAROO or VERITAS collaborations, very high annihilation rates
are required. This, in turn, requires a very large annihilation
 cross-section {\it and} a very concentrated dark matter distribution in the
innermost region of our Galaxy. Even if we consider a particle with a
rather large annihilation cross-section, say $\sim 10^{-26} \,
\rm{cm}^3/\rm{s}$, extremely cusped halo models, such as a Moore {\it
et al.} with adiabatic compression would be required. Alternatively,
halo profiles with a density spike most plausibly associated with
the central SMBH  or a nearby VMBH could provide the observed flux.

\subsection{Future Prospects}

With the current data, it is very difficult to determine whether the
$\gamma$-rays observed from the Galactic center region by CANGAROO-II
and Whipple are the product of dark matter annihilations rather than
other, less exotic, astrophysics. This state of affairs may change
with improved data in the future.
As the angular resolution of ACTs (as well as space-based $\gamma$-ray
experiments, such as GLAST) improves, it will become clear whether the
observed TeV emission comes from our Galaxy's dynamical center rather
than from nearby star clusters, X-ray binaries or other objects. This
information will be crucial for confidently identifying TeV emission
as the product of dark matter annihilations.

Moreover, as the $\gamma$-ray spectrum in the GeV to multi-TeV range becomes
more refined, it may become possible to ascertain whether the observed
emission is the result of annihilating dark matter. In particular, 
evidence of line emission would provide a ``smoking gun'' signal
for anihilations.
Presently, we are eagerly awaiting results from the HESS
collaboration, which has also been observing the Galactic center. HESS
should be more sensitive in the direction of the Galactic center than
either CANGAROO-II or Whipple. Also, with four telescopes, HESS's
angular resolution should be superior to single-telescope ACTs.

\section*{Acknowledgments}

We would like to thank Reba Bandyopadhyay for helpful discussions. DH
and FF are supported by the Leverhulme trust.


\begin{thebibliography}{99}
\singlespaced

\bibitem{Kosack:2004ri} K.~Kosack [the VERITAS Collaboration],
arXiv:astro-ph/0403422.

\bibitem{Tsuchiya:2004wv} K.~Tsuchiya {\it et al.}  [CANGAROO-II
Collaboration], 
Galactic Center Direction by
arXiv:astro-ph/0403592.



\bibitem{Ghez:2003hb} A.~M.~Ghez {\it et al.},
Astrophys.\ J.\ {\bf 601}, L159 (2004) [arXiv:astro-ph/0309076].

\bibitem{xray} F.~K.~Baganoff {\em et al.}, Nature, {\bf 413}, 45
(2001).

\bibitem{Belanger:2003se} G.~Belanger {\it et al.},
Astrophys.\ J.\ {\bf 601}, L163 (2004) [arXiv:astro-ph/0311147].


\bibitem{microquasar_gamma} J.~M.~Paredes, J.~Mart\'\i, M.~Rib\'o and
M.~Massi, Science, 288, 2340 (2000).

\bibitem{microquasar_proton} G.~E.~Romero, D.~F.~Torres, M.~M.~Kaufman
Bernad\'o and I.~F.~Mirabel, A$\&$A 410, L1 (2003).

\bibitem{microquasar_lepton} G.~E.~Romero, M.~M.~Kaufman Bernad\'o and
I.~F.~Mirabel, A$\&$A 393, L61 (2002).


\bibitem{snr_crab} A.~M.~Hillas, {\it et al.}, Astrophys.\ J.\ {\bf
503}, 744 (1998).



\bibitem{snr_casa} F.~A.~Aharonian, {\it et al.}, A$\&$A 370, 112
(2001b).



\bibitem{Volk_snr} H.~J.~V\"{o}lk, Proceedings of ESO-CERN-ESA
Symposium on Astronomy, Cosmology and Fundamental Physics, Garching,
Germany, 4-7 Mar 2002 [arXiv:astro-ph/0210297];
D.~F.~Torres, G.~E.~Romero, T.~M.~Dame, J.~A.~Combi and Y.~M.~Butt,
Phys.\ Rept.\  {\bf 382}, 303 (2003)
[arXiv:astro-ph/0209565].




\bibitem{SgrA_East_SNR} Y.~Maeda, {\it et al.}, Astrophys.\ J.\ {\bf
570}, 671 (2002).



\bibitem{Aharonian94_SNRMOLECULAR} F.~A.~Aharonian, L.~Drury and H.~J.~Volk,
A$\&$A 285, 645 (1994); M.~Fatuzzo and F.~Melia,
Astrophys.\ J.\  {\bf 596}, 1035 (2003)
[arXiv:astro-ph/0302607].



\bibitem{TeVOBObservations} F.~Aharonian, {\it et al.}, 
A$\&$A 393, L37 (2002).

\bibitem{TeVOBTheory} D.~F.~Torres, E.~Domingo-Santamar\'\i a and G.~E.~Romero, 
Astrophys.\ J.\  {\bf 601}, L75 (2004).


\bibitem{YusefArches1} F.~Yusef-Zadeh, {\it et al.},
Astrophys.\ J.\  {\bf 570}, 665 (2002).


\bibitem{YusefArches2} F.~Yusef-Zadeh, {\it et al.},
Astrophys.\ J.\  {\bf 590}, L103 (2003).


\bibitem{INTEGRAL_SOURCES}
G.~Di Cocco, {\it et al.},
arXiv:astro-ph/0403676.


\bibitem{dingus}
D.~Hooper and B.~L.~Dingus,
arXiv:astro-ph/0210617;
D.~Hooper and B.~Dingus,
34th COSPAR Scientific Assembly: The 2nd World Space Congress, Houston, Texas, 10-19 Oct 2002 [arXiv:astro-ph/0212509].


\bibitem{Galactic_XRAY} Q.~D.~Wang, E.~V.~Gotthelf and C.~C.~Lang, 
Nature, 415, 148 (2002). 

\bibitem{YusefZadeh00} F.~Yusef-Zadeh, F.~Melia and M.~Wardle, 
Science 287, 85 (2000).

\bibitem{LaRosa00} T.~N.~LaRosa, N.~E.~Kassim and T.~J.~W.~Lazio,
Astrophys.\ J.\  {\bf 119}, L207 (2000).





\bibitem{cdm}
P.~J.~E.~Peebles,
Astrophys.\ J.\ {\bf 277}, 470 (1984);
C.~L.~Bennett {\it et al.},
Astrophys.\ J.\ Suppl.\  {\bf 148}, 1 (2003)
[arXiv:astro-ph/0302207].

\bibitem{direct}
A.~Drukier and L.~Stodolsky,
Phys.\ Rev.\ D {\bf 30}, 2295 (1984);
M.~W.~Goodman and E.~Witten,
Phys.\ Rev.\ D {\bf 31}, 3059 (1985);
P.~F.~Smith,
Phil.\ Trans.\ Roy.\ Soc.\ Lond.\ A {\bf 361}, 2591 (2003);
H.~Kraus,
Phil.\ Trans.\ Roy.\ Soc.\ Lond.\ A {\bf 361}, 2581 (2003).

\bibitem{indirectneutrino}
J.~Silk, K.~Olive and M.~Srednicki,
Phys.\ Rev.\ Lett.\ {\bf 55}, 257 (1985);
K.~Freese,
Phys.\ Lett.\ B {\bf 167}, 295 (1986);
L.~M.~Krauss, M.~Srednicki and F.~Wilczek,
Phys.\ Rev.\ D {\bf 33}, 2079 (1986);
T.~K.~Gaisser, G.~Steigman and S.~Tilav,
Phys.\ Rev.\ D {\bf 34}, 2206 (1986);
L.~Bergstrom, J.~Edsjo and P.~Gondolo,
Phys.\ Rev.\ D {\bf 58}, 103519 (1998);
D.~Hooper and J.~Silk,
New J.\ Phys.\  {\bf 6}, 023 (2004)
[arXiv:hep-ph/0311367].

\bibitem{positrons}
M.~Kamionkowski and M.~S.~Turner,
Phys.\ Rev.\ D {\bf 43}, 1774 (1991);
E.~A.~Baltz and J.~Edsjo,
Phys.\ Rev.\ D {\bf 59} (1999) 023511
[arXiv:astro-ph/9808243];
G.~L.~Kane, L.~T.~Wang and J.~D.~Wells,
Phys.\ Rev.\ D {\bf 65}, 057701 (2002);
W.~de Boer, M.~Herold, C.~Sander and V.~Zhukov,
arXiv:hep-ph/0309029.
D.~Hooper, J.~E.~Taylor and J.~Silk,
Phys.\ Rev.\ D, in press [arXiv:hep-ph/0312076].

\bibitem{antiprotons}
J.~Silk and M.~Srednicki,
Phys.\ Rev.\ Lett.\  {\bf 53}, 624 (1984);
F.~W.~Stecker, S.~Rudaz and T.~F.~Walsh,
Phys.\ Rev.\ Lett.\  {\bf 55}, 2622 (1985);
L.~Bergstrom, J.~Edsjo and P.~Ullio,
26th International Cosmic Ray Conference (ICRC 99), Salt Lake City, UT, 17-25 Aug 1999, [arXiv:astro-ph/9906034];
F.~Donato, N.~Fornengo, D.~Maurin, P.~Salati and R.~Taillet,
Phys.\ Rev.\ D {\bf 69}, 063501 (2004)
[arXiv:astro-ph/0306207].


\bibitem{indirectgamma}
S.~Rudaz and F.~W.~Stecker,
Astrophys.\ J.\  {\bf 325}, 16 (1988);
H.~U.~Bengtsson, P.~Salati and J.~Silk,
Nucl.\ Phys.\ B {\bf 346}, 129 (1990);
V.~Berezinsky, A.~Bottino and G.~Mignola,
Phys.\ Lett.\ B {\bf 325}, 136 (1994)
[arXiv:hep-ph/9402215];
F.~Stoehr, S.~D.~M.~White, V.~Springel, G.~Tormen and N.~Yoshida,
Mon.\ Not.\ Roy.\ Astron.\ Soc.\  {\bf 345}, 1313 (2003)
[arXiv:astro-ph/0307026];
N.~W.~Evans, F.~Ferrer and S.~Sarkar,
Phys.\ Rev.\ D, in press,
arXiv:astro-ph/0311145.

\bibitem{Bergstrom:1997fj}
L.~Bergstrom, P.~Ullio and J.~H.~Buckley,
Astropart.\ Phys.\  {\bf 9}, 137 (1998)
[arXiv:astro-ph/9712318].

\bibitem{Bergstrom:2001jj}
L.~Bergstrom, J.~Edsjo and P.~Ullio,
Phys.\ Rev.\ Lett.\  {\bf 87}, 251301 (2001)
[arXiv:astro-ph/0105048].

\bibitem{Thompson93} 
D.~J.~Thompson, {\it et al},
Astrophys.\ J.\ Suppl.\  {\bf 86}, 629 (1993).


\bibitem{Gehrels:ri}
N.~Gehrels and P.~Michelson,
Astropart.\ Phys.\  {\bf 11} (1999) 277.

\bibitem{Hillas96} 
A.\ M.\ Hillas,
Space Science Reviews {\bf 75}, 17 (1996).

\bibitem{Reynolds:av}
P.~T.~Reynolds {\it et al.},
Astrophys.\ J.\  {\bf 404}, 206 (1993).

\bibitem{whipple} 
M. F. Cawley {\em et al.},
Experimental Astronomy {\bf 1}, 173 (1990);
J. P. Finley {\em et al.}, 
Proc. 27th ICRC, Hamburg, {\bf 7}, 2827 (2001).

\bibitem{Kawachi01} 
A.~Kawachi {\em et al.},
Astroparticle Physics, {\bf 14}, 261 (2001);
K.~Tsuchiya and R.~Enomoto  [CANGAROO Collaboration],
Prepared for International Symposium: The Universe Viewed in Gamma Rays,
Kashiwa, Chiba, Japan, 25-28 Sep 2002;
K.~Tsuchiya, R.~Enomoto and M.~Mori  [CANGAROO Collaboration],
Prepared for 28th International Cosmic Ray Conferences (ICRC 2003), 
Tsukuba, Japan, 31 Jul - 7 Aug 2003.

\bibitem{pythia}
T.~Sjostrand, P.~Eden, C.~Friberg, L.~Lonnblad, G.~Miu, S.~Mrenna and E.~Norrbin,
Comput.\ Phys.\ Commun.\  {\bf 135}, 238 (2001)
[arXiv:hep-ph/0010017].

\bibitem{darksusy}
P.~Gondolo, J.~Edsjo, L.~Bergstrom, P.~Ullio and E.~A.~Baltz,
arXiv:astro-ph/0012234.


\bibitem{Birkel:1998nx}
M.~Birkel and S.~Sarkar,
Astropart.\ Phys.\  {\bf 9}, 297 (1998)
[arXiv:hep-ph/9804285].

\bibitem{Sarkar:2001se}
S.~Sarkar and R.~Toldra,
Nucl.\ Phys.\ B {\bf 621}, 495 (2002)
[arXiv:hep-ph/0108098].

\bibitem{loop}
P.~Ullio and L.~Bergstrom,
Phys.\ Rev.\ D {\bf 57}, 1962 (1998)
[arXiv:hep-ph/9707333];
L.~Bergstrom and P.~Ullio,
Nucl.\ Phys.\ B {\bf 504}, 27 (1997)
[arXiv:hep-ph/9706232].

\bibitem{nfw}
J.~F.~Navarro, C.~S.~Frenk and S.~D.~White,
Astrophys.\ J.\  {\bf 462}, 563 (1996)
[arXiv:astro-ph/9508025];
J.~F.~Navarro, C.~S.~Frenk and S.~D.~White,
Astrophys.\ J.\  {\bf 490}, 493 (1997).

\bibitem{moore}
B.~Moore, S.~Ghigna, F.~Governato, G.~Lake, T.~Quinn, J.~Stadel and
P.~Tozzi,
Astrophys.\ J.\  {\bf 524}, L19 (1999).

\bibitem{nbody}
J.~F.~Navarro, C.~S.~Frenk and S.~D.~M.~White,
Astrophys.\ J.\  {\bf 462}, 563 (1996)
[arXiv:astro-ph/9508025];
J.~F.~Navarro, C.~S.~Frenk and S.~D.~M.~White,
Astrophys.\ J.\  {\bf 490}, 493 (1997);
B.~Moore, S.~Ghigna, F.~Governato, G.~Lake, T.~Quinn, J.~Stadel and P.~Tozzi,
Astrophys.\ J.\  {\bf 524}, L19 (1999).

\bibitem{Binney:2003sn}
J.~Binney,
arXiv:astro-ph/0311155;
J.~F.~Navarro {\it et al.},
arXiv:astro-ph/0311231.





\bibitem{klypin}
F.~Prada, A.~Klypin, J.~Flix, M.~Martinez and E.~Simonneau,
arXiv:astro-ph/0401512.

\bibitem{spike}
P.~Gondolo and J.~Silk, 
Phys.\ Rev.\ Lett.\ {\bf 83}, 1719 (1999) 
[arXiv:astro-ph/9906391];  
P.~Ullio, H.~Zhao and M.~Kamionkowski, 
Phys.\ Rev.\ D {\bf 64}, 043504 (2001)  
[arXiv:astro-ph/0101481]; 
G.~Bertone, G.~Sigl and J.~Silk, 
Mon.\ Not.\ Roy.\ Astron.\ Soc.\  {\bf 337}, 98 (2002)
[arXiv:astro-ph/0203488].

\bibitem{merrit}
D.~Merritt, M.~Milosavljevic, L.~Verde and R.~Jimenez,
arXiv:astro-ph/0201376.

\bibitem{merritta}
D.~Merritt,
astro-ph/0311594

\bibitem{bertone}
G.~Bertone, G.~Sigl and J.~Silk, 
Mon.Not.Roy.Astron.Soc. 326 (2001) 799-804

\bibitem{abo}
R.~Aloisio, P.~Blasi and A.~V.~Olinto,
arXiv:astro-ph/0402588.

\bibitem{milos}
M.~Milosavljevic and D.~Merritt,
Review, to appear in: ``The Astrophysics of Gravitational Wave
Sources'', 
J. Centrella (ed.), AIP, in press (2003)
[arXiv:astro-ph/0212270]

\bibitem{madau}
M.~Volonteri, F.~Hardt and P.~Madau, 
Astrophys.J. 582 (2003) 559-573

\bibitem{islam}
R.~Islam, J.~Taylor and J.~Silk, 
astro-ph/0307171, MNRAS, in press (2004)

\bibitem{be}
J.~J.~Binney and N.~W.~Evans,
Mon.\ Not.\ Roy.\ Astron.\ Soc.\  {\bf 327}, L27 (2001)
[arXiv:astro-ph/0108505].

\bibitem{kzs}
A.~Klypin, H.~Zhao and R.~S.~Somerville,
arXiv:astro-ph/0110390.


\bibitem{jackson}
J.~M.~Jackson, {\it et al.},
Astrophys.\ J.\  {\bf 456}, L91 (1996).

\bibitem{focusnatural}
J.~L.~Feng, K.~T.~Matchev and F.~Wilczek,
Phys.\ Lett.\ B {\bf 482}, 388 (2000)
[arXiv:hep-ph/0004043];
J.~L.~Feng, K.~T.~Matchev and T.~Moroi,
Phys.\ Rev.\ Lett.\  {\bf 84}, 2322 (2000)
[arXiv:hep-ph/9908309];
J.~L.~Feng, K.~T.~Matchev and T.~Moroi,
Phys.\ Rev.\ D {\bf 61}, 075005 (2000)
[arXiv:hep-ph/9909334];
M.~Drees, M.~M.~Nojiri, D.~P.~Roy and Y.~Yamada,
Phys.\ Rev.\ D {\bf 56}, 276 (1997)
[Erratum-ibid.\ D {\bf 64}, 039901 (2001)]
[arXiv:hep-ph/9701219].



\bibitem{resonances}
M.~Drees and M.~M.~Nojiri,
Phys.\ Rev.\ D {\bf 47}, 376 (1993)
[arXiv:hep-ph/9207234].


\bibitem{coannihilations}
C.~Boehm, A.~Djouadi and M.~Drees,
Phys.\ Rev.\ D {\bf 62}, 035012 (2000)
[arXiv:hep-ph/9911496];
J.~R.~Ellis, T.~Falk, K.~A.~Olive and M.~Srednicki,
Astropart.\ Phys.\  {\bf 13}, 181 (2000)
[Erratum-ibid.\  {\bf 15}, 413 (2001)]
[arXiv:hep-ph/9905481].



\bibitem{amsb}
D.~Majumdar,
J.\ Phys.\ G {\bf 28}, 2747 (2002)
[arXiv:hep-ph/0209278];
P.~Ullio,
JHEP {\bf 0106}, 053 (2001)
[arXiv:hep-ph/0105052];
D.~Hooper and L.~T.~Wang,
Phys.\ Rev.\ D {\bf 69}, 035001 (2004)
[arXiv:hep-ph/0309036].

\bibitem{gammakk}
G.~Servant and T.~M.~P.~Tait,
Nucl.\ Phys.\ B {\bf 650}, 391 (2003)
[arXiv:hep-ph/0206071];
H.~C.~Cheng, J.~L.~Feng and K.~T.~Matchev,
Phys.\ Rev.\ Lett.\  {\bf 89}, 211301 (2002)
[arXiv:hep-ph/0207125];
G.~Bertone, G.~Servant and G.~Sigl,
Phys.\ Rev.\ D {\bf 68}, 044008 (2003)
[arXiv:hep-ph/0211342].

\bibitem{lhdm}
A.~Birkedal-Hansen and J.~G.~Wacker,
arXiv:hep-ph/0306161.

\bibitem{heavynu}
K.~Enqvist, K.~Kainulainen and J.~Maalampi,
Nucl.\ Phys.\ B {\bf 317}, 647 (1989);
P.~Roy,
ICNAPP 1994:0225-237, [arXiv:hep-ph/9501209].


\end{thebibliography}
\end{document}